\documentclass[12pt, twoside, a4paper]{article}
\usepackage{imm17_4}

\pdfoutput=1

\usepackage[pdftex, unicode]{hyperref}
\usepackage[pdftex]{graphicx}

\hypersetup{pdftitle={Generalized synchronization of multidimensional chaotic systems in terms of symbolic CTQ-analysis}}
\hypersetup{pdfauthor={A. V. Makarenko, Constructive Cybernetics Research Group}}
\hypersetup{pdfsubject={Nonlinear Dynamics}}
\hypersetup{pdfkeywords={Chaotic systems, Generalized synchronization, Attractor's structure, Intermittency of synchronism, Symbolic CTQ-analysis}}

\newcommand{\MaincltStr}{arXiv.org}

\newcommand{\Leftclt}{A.\,V.\,Makarenko}

\newcommand{\Rightclt}{Generalized synchronization of multidimensional chaotic systems\ldots}

\newcommand{\UDK}{MSC: 37M10, 37M20, 34C28, 34C15, 34D06;
\\ PACS: 02.70.-c, 05.45.-a, 05.45.Tp, 05.45.Xt.}

\begin{document}

\Mainclt 

\begin{center}
\Large{\bf Generalized synchronization\\ of multidimensional chaotic systems\\ in terms of symbolic CTQ-analysis}\\[2ex]
\end{center}

\begin{center}
\large\bf{A.\,V.\,Makarenko}\supit{a,}\supit{b,}
\footnote{E-mail: avm.science@mail.ru}\\[2ex]
\end{center}

\begin{center}
\supit{a}\normalsize{Constructive Cybernetics Research Group}
\\
\normalsize{P.O.Box~560, Moscow, 101000 Russia}\\[3ex]

\supit{b}
\normalsize{Institute of Control Sciences, Russian Academy of Sciences}
\\
\normalsize{ul.~Profsoyuznaya~65, Moscow, 117977 Russia}\\[3ex]
\end{center}

\begin{quote}\small
{\bf Abstract}. A new approach is proposed to the analysis of generalized synchronization of multidimensional chaotic systems. The approach is based on the symbolic analysis of discrete sequences in the basis of a finite T-alphabet. In fact, the symbols of the T-alphabet encode the shape (the geometric structure) of a trajectory of a dynamical system. Investigation of symbolic sequences allows one to diagnose various regimes of chaos synchronization, including generalized synchronization. The characteristics introduced allow one to detect and study the restructuring and intermittency behavior of attractors in systems (the time structure of synchronization). The measure of T-synchronization proposed is generalized without restrictions to complex ensembles of strongly nonstationary and nonidentical large-dimensional oscillators with arbitrary configuration and network (lattice) topology. The main features of the method are illustrated by an example.
\end{quote}

\begin{Keyworden}
Chaotic systems, Generalized synchronization, Attractor's structure, Intermittency of synchronism, Symbolic CTQ-analysis.
\end{Keyworden}


\setcounter{equation}{0}
\setcounter{lem}{0}
\setcounter{teo}{0}



\section{Introduction}

Synchronization is one of the fundamental concepts of the theory of nonlinear dynamics and chaos
theory. This phenomenon is widespread in nature, science, engineering, and society [1]. One of important
manifestations of this phenomenon is the synchronization of chaotic oscillations, which was experimentally observed in various physical applications (see~\cite{bib:book_Pikovsky_2001, bib:article_Boccaletti_PhysRep_2002_366, bib:article_Argonov_JETPL_2004_80, bib:article_Kuznetsov_UFN_2011_2, bib:article_Napartovich_JETP_1999_5} and references therein) such as radio oscillators, mechanical systems, lasers, electrochemical oscillators, plasma and gas discharge, and quantum systems. The study of this phenomenon is also very important from the viewpoint of its application to information transmission~\cite{bib:article_Cuomo_PhysRevLett_1993_71}, cryptographic coding~\cite{bib:article_Larger_Physique_2004_5} with the use of deterministic chaotic oscillations, and quantum computation~\cite{bib:article_Argonov_JETPL_2004_80, bib:article_Planat_Neuroquantology_2004_2}.

There are several types of synchronization of chaotic oscillations~\cite{bib:article_Boccaletti_PhysRep_2002_366}:
generalized synchronization~\cite{bib:article_Abarbanel_PhysRevE_1996_53},
complete synchronization~\cite{bib:article_Pecora_PhysRevLett_1990_64},
antisynchronization~\cite{bib:article_Liu_PhysLettA_2006_354},
lag synchronization~\cite{bib:article_Rosenblum_PhysRevLett_1997_78},
frequency synchronization~\cite{bib:article_Anishenko_TechPhysLett_1988_6},
phase synchronization~\cite{bib:article_Pikovsky_JourBifChaos_2000_10},
time scale synchronization~\cite{bib:article_Koronovskii_JETPL_2004_79},
and T-synchronization~\cite{bib:article_Makarenko_JETP_2015_5}.
For each type, an appropriate analytic apparatus and diagnostic methods have been developed.
Nevertheless, intensive investigations are being continued that are aimed, on the one hand,
at the examination of different types of synchronization from unique positions and, on the
other hand, at the search for new types of synchronous behavior that do not fall under the
above-mentioned types. In spite of the long history of the study of synchronization of chaotic
oscillations, many important problems in this field remain unsolved.

These include generalized synchronization in the form
\begin{equation}\label{eq:gen_sync}
\mathbf{y}(t)=\mathbf{F}\bigl[{\mathbf{x}(t),\,\boldsymbol{\tau}}\bigr],
\end{equation}
where~$\mathbf{x}$ and~$\mathbf{y}$ are multidimensional
synchronized systems, $\mathbf{F}$ is a function of generalized
link between the systems, and $\boldsymbol{\tau}$ is a delay
vector between the phase variables of the systems~$\mathbf{x}$
and~$\mathbf{y}$.

In this paper, we develop an original method for the diagnostics and quantitative
measurement of the characteristics of generalized synchronization of chaotic systems,
which is aimed at the integrated study of the time structure of synchronism through the
analysis of the so-called T-synchronization~\cite{bib:article_Makarenko_TechPhysLett_2012_9, bib:article_Makarenko_JETP_2015_5}.

The method is based on the formalism of symbolic CTQ-analysis
proposed by the present
author~\cite{bib:article_Makarenko_TechPhysLett_2012_38_155,
bib:article_Makarenko_ComputMathMathPhys_2012_52_1017} (the
abbreviation CTQ stands for three alphabets with which the method
operates: C, T, and Q). One should note that symbolic dynamics,
for all of its seeming external simplicity, is a very strongly
substantiated tool for the analysis of nonlinear dynamical
systems~\cite{bib:book_Gilmore_2002, bib:book_Guckenheimer_1997,
bib:article_Bowen_AJM_1973_95}.

This article is an expanded version of the report~\cite{bib:report_Makarenko_CHAOS_2015}.

\section{The Symbolic CTQ-analysis}

Denote a discrete dynamical system as a mapping
\begin{equation}\label{eq:ds}
\mathbf{s}_{k+1} =
\mathbf{f}\left(\mathbf{s}_{k},\,\mathbf{p}\right)
\end{equation}
with the following properties:
 $\mathbf{s}\in\mathrm{S}\subseteq\mathbb{R}^N$, $k\in\mathrm{K}\subseteq\mathbb{Z}$,
 $\mathbf{p}\in\mathrm{P}\subseteq\mathbb{R}^M$, $n\in\overline{1,\,N}$, $m\in\overline{1,\,M}$.
In formula~(\ref{eq:ds}), $\mathbf{s}$ is a state variable of the
system and $\mathbf{p}$ is a vector of parameters. With
mapping~(\ref{eq:ds}), we associate its trajectory in the
space~$\mathrm{S}\times\mathrm{K}$, which has the form of a
semisequence~$\{\mathbf{s}_{k}\}^K_{k=1}$, $k\in\overline{1,\,K}$.

\subsection{T-alphabet}

Define the initial mapping, which encodes (in terms of the final
T-alphabet) the shape of the $n$-th component of the
sequence~$\{\mathbf{s}_{k}\}^K_{k=1}$~\cite{bib:article_Makarenko_TechPhysLett_2012_38_155,
bib:article_Makarenko_ComputMathMathPhys_2012_52_1017}:
\begin{equation}\label{eq:mapping_TSymb}
\left\lbrace\mathbf{s}^{(n)}_{k-1},\,\mathbf{s}^{(n)}_{k},\,\mathbf{s}^{(n)}_{k+1}\right\rbrace\Rightarrow
T^{\alpha\varphi}_k|_n,
\quad
T^{\alpha\varphi}_k = \left[
T^{\alpha\varphi}_k|_1,\,\ldots,\,
T^{\alpha\varphi}_k|_n,\,\ldots,\,
T^{\alpha\varphi}_k|_N\right].
\end{equation}

\begin{figure}
\centering
\begin{tabular}{cc}
\begin{minipage}{170pt}
\includegraphics[width=168pt]{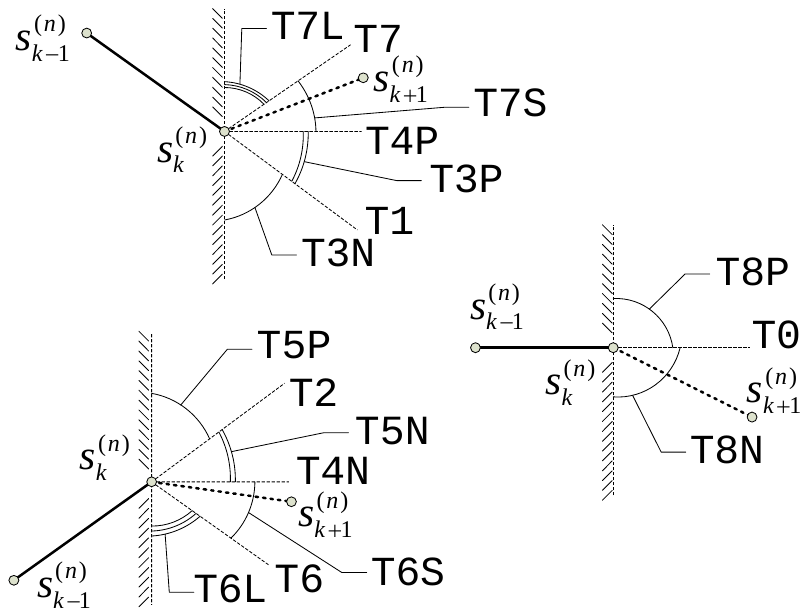}
\caption{Geometry of T-alphabet symbols.}
\label{fig:TSymb_diag}
\end{minipage}
&
\centering
\begin{minipage}{170pt}
\includegraphics[width=168pt]{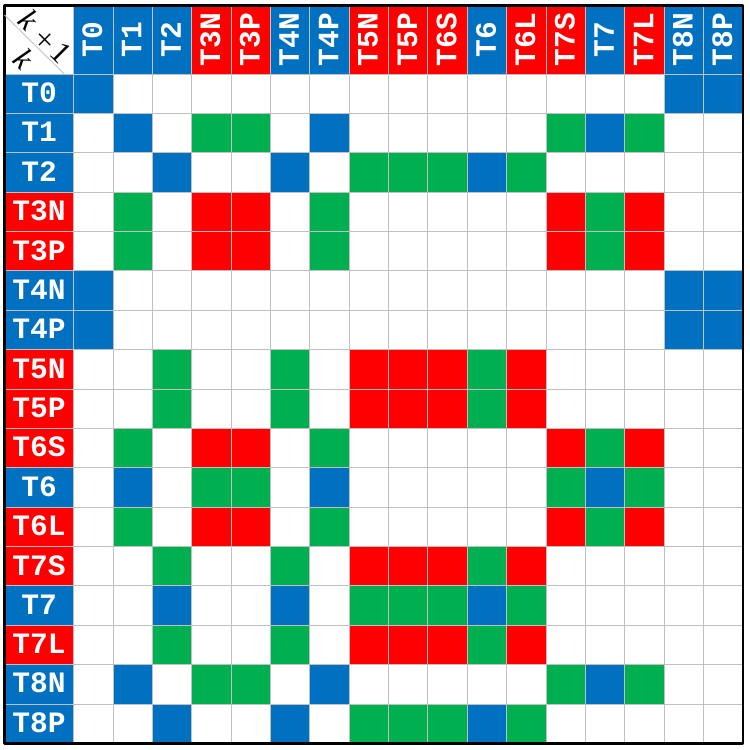}
\caption{Transition matrix of T-alphabet symbols.}
\label{fig:TSymb_trans}
\end{minipage}
\end{tabular}
\end{figure}

The graphic diagrams illustrating the geometry of the symbols~$T^{\alpha\varphi}|_n$ for
the~$k$-th sample and the~$n$-th phase variable are shown in Figure~\ref{fig:TSymb_diag}.

Strictly speaking, the mapping~(\ref{eq:mapping_TSymb}) is defined
by the relations:
\begin{equation}\label{eq:TSymb}
\begin{aligned}
 &\mathtt{T0}\quad   &&\Delta s_-=\Delta s_+=0,     \\
 &\mathtt{T1}\quad   &&\Delta s_-=\Delta s_+<0,     \\
 &\mathtt{T2}\quad   &&\Delta s_-=\Delta s_+>0,     \\
 &\mathtt{T3N}\quad  &&\Delta s_-<0,\quad \Delta s_+<\Delta s_-,  \\
 &\mathtt{T3P}\quad  &&\Delta s_-<0,\quad \Delta s_+<0,\quad \Delta s_+>\Delta s_-,  \\
 &\mathtt{T4N}\quad  &&\Delta s_->0,\quad \Delta s_+=0,  \\
 &\mathtt{T4P}\quad  &&\Delta s_-<0,\quad \Delta s_+=0,  \\
 &\mathtt{T5N}\quad  &&\Delta s_->0,\quad \Delta s_+>0,\quad \Delta s_+<\Delta s_-,  \\
 &\mathtt{T5P}\quad  &&\Delta s_->0,\quad \Delta s_+>\Delta s_-,  \\
 &\mathtt{T6S}\quad  &&\Delta s_->0,\quad \Delta s_+<0,\quad \Delta s_+> -\Delta s_-,  \\
 &\mathtt{T6}\quad   &&\Delta s_-=-\Delta s_+>0,  \\
 &\mathtt{T6L}\quad  &&\Delta s_->0,\quad \Delta s_+<0,\quad \Delta s_+< -\Delta s_-,  \\
 &\mathtt{T7S}\quad  &&\Delta s_-<0,\quad \Delta s_+>0,\quad \Delta s_+< -\Delta s_-,  \\
 &\mathtt{T7}\quad   &&\Delta s_-=-\Delta s_+<0,  \\
 &\mathtt{T7L}\quad  &&\Delta s_-<0,\quad \Delta s_+>0,\quad \Delta s_+> -\Delta s_-,  \\
 &\mathtt{T8N}\quad  &&\Delta s_-=0,\quad \Delta s_+<0,  \\
 &\mathtt{T8P}\quad  &&\Delta s_-=0,\quad \Delta s_+>0.
\end{aligned}\,,
\end{equation}
Here~$\Delta s_- = \mathbf{s}^{(n)}_{k}-\mathbf{s}^{(n)}_{k-1}$
and~$\Delta s_+ = \mathbf{s}^{(n)}_{k+1}-\mathbf{s}^{(n)}_{k}$.

Thus, the T-alphabet includes the following set of symbols:
\begin{multline} \label{eq:T_alphabet_o_n}
\mathrm{T}^{\alpha\varphi}_o=\{\mathtt{T0},\,\mathtt{T1},\,\mathtt{T2},\,\mathtt{T3N},
\,\mathtt{T3P},\,\mathtt{T4N},\,\mathtt{T4P},\,\mathtt{T5N},\,\mathtt{T5P},\,
\\
\mathtt{T6S},\,\mathtt{T6},\,\mathtt{T6L},\,\mathtt{T7S},\,\mathtt{T7},\,\mathtt{T7L},
\,\mathtt{T8N},\,\mathtt{T8P}\}.
\end{multline}

One can see from~(\ref{eq:T_alphabet_o_n}) that the symbol~$T^{\alpha\varphi}_k|_n$ is encoded
as~$\mathtt{T}\,i$, where~$i$ is the right-hand side of the symbol codes of the
alphabet~$\mathrm{T}^{\alpha\varphi}_o$. In turn, the symbol~$T^{\alpha\varphi}_k$ is encoded
in terms of~$\mathtt{T}\,i_1\,\cdots\,i_n\,\cdots\,i_N$, see~(\ref{eq:mapping_TSymb}). The full
alphabet~$\mathrm{T}^{\alpha\varphi}_o|N$, which encodes the shape of the trajectory of the
multidimensional sequence~$\{\mathbf{s}_k\}^K_{k=1}$, consists of~$17^N$~symbols.

\subsection{Q-alphabet}

In addition to the symbols~$T^{\alpha\varphi}_k|_n$, we introduce the
symbols~$Q^{\alpha\varphi}_k|_n$:
\begin{equation}\label{eq:SymbQ}
Q^{\alpha\varphi}_k|_n\equiv T^{\alpha\varphi}_k|_n\rightarrow T^{\alpha\varphi}_{k+1}|_n,
\quad
Q^{\alpha\varphi}_k = \left[
Q^{\alpha\varphi}_k|_1,\,\ldots,\,
Q^{\alpha\varphi}_k|_n,\,\ldots,\,
Q^{\alpha\varphi}_k|_N\right].
\end{equation}

All admissible transitions constitute a set of symbols of the
alphabet~$\mathrm{Q}^{\alpha\varphi}_o\ni Q^{\alpha\varphi}_k|_n$. These transitions are shown
in Figure~\ref{fig:TSymb_trans}.

The symbol~$Q^{\alpha\varphi}_k|_n$ is encoded
as~$\mathtt{Q}\,i\,j$, where~$i$ and~$j$ are the right-hand sides
of the symbol codes of the alphabet~$\mathrm{T}^{\alpha\varphi}_o$
for the states~$k$ and~$k+1$, respectively. In turn, the
symbol~$Q^{\alpha\varphi}_k$ is encoded in terms
of~$\mathtt{Q}\,i_1\,\cdots\,i_n\,\cdots\,i_N\,j_1\,\cdots\,j_n\,\cdots\,j_N$,
see~(\ref{eq:SymbQ}). The full
alphabet~$\mathrm{Q}^{\alpha\varphi}_o|N$, which encodes the shape
of the trajectory of the sequence~$\{\mathbf{s}_k\}^K_{k=1}$,
consists of~$107^N$~symbols (see Figure~\ref{fig:TSymb_trans}).

\subsection{Symbolic TQ-image of a dynamical system}

Let us introduce a directed graph
\begin{equation}\label{eq:Gamma TQ}
\Gamma^{TQ}=\left\langle\mathrm{V^\Gamma},\,\mathrm{E^\Gamma}\right\rangle,\quad
\mathrm{V^\Gamma}\subseteq\mathrm{T}^{\alpha\varphi}_o|N,\quad
\mathrm{E^\Gamma}\subseteq\mathrm{Q}^{\alpha\varphi}_o|N,
\end{equation}
which is a complete symbolic TQ-image of the dynamical
system~$(\ref{eq:ds})$. By definition,~$\mathrm{V}^{\Gamma}$ is
the vertex set and~$\mathrm{E}^{\Gamma}$ is the edge set
of~$\Gamma^{TQ}$. According to its topology, the
graph~$\Gamma^{TQ}$ has no multiple arcs but has loops.

By analogy with~(\ref{eq:Gamma TQ}), introduce a directed graph
\begin{equation}\label{eq:Gamma_TQ_n}
\Gamma^{TQ}|_n=\left\langle\mathrm{V^\Gamma}|_n,\,\mathrm{E^\Gamma}|_n\right\rangle,\quad
\mathrm{V^\Gamma}|_n\subseteq\mathrm{T}^{\alpha\varphi}_o,\quad
\mathrm{E^\Gamma}|_n\subseteq\mathrm{Q}^{\alpha\varphi}_o,
\end{equation}
which is a particular symbolic TQ-image of the dynamical system
with respect to its $n$-th phase variable. Denote the
graph~$\Gamma^{TQ}|_n$ corresponding to the full
alphabets~$\mathrm{T}^{\alpha\varphi}_o$
and~$\mathrm{Q}^{\alpha\varphi}_o$ by~$\Gamma^{TQ}_o$.

A particular symbolic TQ-image~(\ref{eq:Gamma_TQ_n}) can be
obtained from the graph~$\Gamma^{TQ}$ by the gluing (or
identification) of its vertices, redirection of the edges incident
to these vertices, and the removal of the arising multiple edges.

Introduce the operation of gluing:
\begin{equation}\label{eq:R_TQ_n}
\Gamma^{TQ}|_n = \mathcal{R}^\mathrm{TQ}\Bigl(\Gamma^{TQ},\,n\Bigr).
\end{equation}
Let~$v\in\mathrm{V^\Gamma}$, $e\in\mathrm{V^\Gamma}$,
$v'\in\mathrm{V^\Gamma}|_n$, and~$e'\in\mathrm{V^\Gamma}|_n$. Then
\begin{equation}\label{eq:Gamma_Symb}
\begin{aligned}
 &v'\equiv\mathtt{T}\,i, \quad &&v\equiv\mathtt{T}\,i_1\,\cdots\,i_n\,\cdots\,i_N,\\
 &e'\equiv\mathtt{Q}\,i\,j, \quad && e\equiv \mathtt{Q}\,i_1\,\cdots\,i_n\,\cdots\,i_N\,j_1\,\cdots\,j_n\,\cdots\,j_N.
\end{aligned}
\end{equation}
The gluing of vertices and edges is formally expressed by the
following conditions:
\begin{equation}\label{eq:R_TQ_n_formaly}
\begin{aligned}
&\mathrm{V^\Gamma}|_n\ni\mathtt{T}\,j:\,
\mathtt{T}\,i_1\,\cdots\,i_n\,\cdots\,i_N\in\mathrm{V^\Gamma},\,i_n=j,\\
&\mathrm{E^\Gamma}|_n\ni\mathtt{Q}\,i\,j:\,
\mathtt{Q}\,k_1\,\cdots\,k_n\,\cdots\,k_N\,l_1\,\cdots\,l_n\,\cdots\,l_N\in\mathrm{E^\Gamma},
\,k_n=i,\,l_n=j.
\end{aligned}
\end{equation}

The graph~(\ref{eq:Gamma_TQ_n}) can be weighted (on its vertices and edges) by the occurrence
frequency of characters~$*$ in the sequence~$\{\mathbf{s}^{(n)}_{k}\}^K_{k=1}$:
\begin{equation}\label{eq:Delta_o}
\Delta^*|_n = \frac{\left| {\mathrm{M}^*|_n} \right|}{
\left|\bigcup\limits_{*}\mathrm{M}^*|_n\right|},\quad 0\leqslant\Delta^*|_n\leqslant 1,
\end{equation}
where $|\cdot|$ is the cardinality of the set and $*$ is a symbol of which the
multiset~$\mathrm{M}^*|_n$ consists:
\begin{subequations}\label{eq:Delta_Symb_All}
\begin{align}
&\Delta^\mathrm{T}|_n: \mathrm{M}^*|_n \ni T^{\alpha\varphi}_k|_n:\,T^{\alpha\varphi}_k|_n\backslash\mathtt{T} = * ,\;
* \in \mathrm{T}^{\alpha\varphi}_o\backslash\mathtt{T},
                \label{eq:Delta_Symb_T} \\[3pt]
&\Delta^\mathrm{Q}|_n: \mathrm{M}^*|_n \ni Q^{\alpha\varphi}_k|_n:\,Q^{\alpha\varphi}_k|_n\backslash\mathtt{Q} = * ,\;
* \in \mathrm{Q}^{\alpha\varphi}_o\backslash\mathtt{Q}.
                \label{eq:Delta_Symb_Q}
\end{align}
\end{subequations}
Similar characteristics~$\Delta^\mathrm{T}$
and~$\Delta^\mathrm{Q}$ are determined for the
graph~$\Gamma^{TQ}$. Note that the calculation of~$\Delta^\mathrm{T}$ and~$\Delta^\mathrm{Q}$ allows one to
quantitatively assess various properties of the trajectory of the sequence~$\{\mathbf{s}_{k}\}^K_{k=1}$
in the space~$\mathrm{S}\times\mathrm{K}$, including the Markov characteristic of the
sequence~$\{T^{\alpha\varphi}_k\}^K_{k=1}$~\cite{bib:book_Gilmore_2002, bib:article_Bowen_AJM_1973_95}.

\section{T-synchronization}

Let us make a remark. For simplicity, but
without loss of generality, suppose that the time
sequence~$\{\mathbf{s}_{k}\}^K_{k=1}$ of dimension~$N$ is formed
by a combination of the phase variables of~$N$ one-dimensional
dynamical systems; i.e., suppose that~$\mathbf{s}^{(n)}_k$ is the
value of the phase variable of the~$n$th system at the~$k$th
instant of time.

\subsection{Complete T-synchronization}
\label{sect:Full_T_Sync}

\textbf{Definition.} Dynamical systems are
\textit{synchronous} at time instant~$k$ in the sense of Complete
T-synchronization~\cite{bib:article_Makarenko_TechPhysLett_2012_9}
if the condition~$J_k=1$ is satisfied, where
\begin{equation} \label{eq:full_J_T_sync}
J_k=\begin{cases}
1 & T^{\alpha\varphi}_k|_1=\ldots=T^{\alpha\varphi}_k|_n=\ldots=T^{\alpha\varphi}_k|_N,\\
0 & \text{otherwise}.
\end{cases}\;.
\end{equation}
Thus,~$\{J_k\}^K_{k=1}$ is the indicator sequence of T-synchronization.

Taking into account possible
antisynchronization~\cite{bib:article_Liu_PhysLettA_2006_354}
between the systems, we should also consider all possible variants
of inversion of their phase variables: $\mathbf{s}^{(n)}_k\to -1
\cdot \mathbf{s}^{(n)}_k$. In this case, for the $n$th system, a
change of symbols~$T^{\alpha\varphi}_k|_n$ in the $k$th sample
occurs according to the scheme
\begin{equation} \label{eq:InvTSymb}
\begin{aligned}
&\mathtt{T0} \leftrightarrow\mathtt{T0},\\[3pt]
&\mathtt{T1} \leftrightarrow\mathtt{T2},\quad
 \mathtt{T3N}\leftrightarrow\mathtt{T5P},\quad
 \mathtt{T3P}\leftrightarrow\mathtt{T5N},\quad
 \mathtt{T4N}\leftrightarrow\mathtt{T4P},\\[3pt]
&\mathtt{T6S}\leftrightarrow\mathtt{T7S},\quad
 \mathtt{T6} \leftrightarrow\mathtt{T7},\quad
 \mathtt{T6L}\leftrightarrow\mathtt{T7L},\quad
 \mathtt{T8N}\leftrightarrow\mathtt{T8P}.
\end{aligned}
\end{equation}
Denote the variants of inversion by number~$m$. The total number
of variants of inversion is~$M=2^{N-1}$.

Synchronization between systems can also be set in the lag
regime~\cite{bib:article_Rosenblum_PhysRevLett_1997_78}. To detect
this synchronization, one should move a little the phase
trajectories of the systems with respect to each other (the shifts
are~$h_n\geqslant 0$):
\begin{equation} \label{eq:Shift_n}
\left\{
T^{\alpha\varphi}_k|_1 \to T^{\alpha\varphi}_{k+h_1}|_1,\, \ldots,\,
T^{\alpha\varphi}_k|_n \to T^{\alpha\varphi}_{k+h_n}|_n,\, \ldots,\,
T^{\alpha\varphi}_k|_N \to T^{\alpha\varphi}_{k+h_N}|_N
\right\}.
\end{equation}

The antisynchronization and lag synchronization regimes may coexist; therefore, when
calculating a partial integral coefficient of synchronism, we take this fact into consideration:
\begin{equation} \label{eq:sigma_sync_h}
\delta^s_{m,\mathbf{h}} = \frac{1}{K^* + 1 - k^*}\sum\limits^{K^*}_{k = k^*} {J_k|\left\{m,\,\mathbf{h}\right\}},
\end{equation}
where $k^* = 1 + \max \left( {h_1,\, \ldots ,\,h_N} \right)$, $K^* = K + \min \left( {h_1,\, \ldots ,\,h_N} \right)$, and~$K$ is the length of the sequence~$\{T^{\alpha\varphi}_k\}^K_{k=1}$.

On the basis of the partial coefficient, we calculate the total integral coefficient of
synchronism of the systems:
\begin{equation} \label{eq:sigma_sync}
\delta^s = \mathop{\max}_m \mathop{\max}_\mathbf{h} \delta^s_{m,\mathbf{h}},\quad
0\leqslant\delta^s\leqslant 1,
\end{equation}
i.e., we take a combination of shifts between the trajectories of
the systems and a variant of inversion of their phase variables
that, taken together, provide the maximum number of samples~$k$
satisfying the condition~$J_k=1$.

\section{Time structure of synchronization of chaotic systems}

The quantity~$\delta^s$ introduced in~(\ref{eq:sigma_sync})
characterizes the synchronism of the systems on average over a
period of~$t_K-t_1$. As mentioned in the Introduction, most
investigations on the synchronization of chaos are usually
restricted to this situation. However, often a researcher may be
interested in the time structure of synchronization of systems.
Recall that by this structure one means the spikes in the
synchronous behavior of the phase variables of the systems between
which the synchronism level is characterized by a small quantity,
i.e., one means intermittent
behavior~\cite{bib:article_Zeldovich_UFN_1987_5,
bib:article_Mandelbrot_JFluidMech_1974_2}.

In~\cite{bib:article_Makarenko_TechPhysLett_2012_9}, the present
author introduced the concept of a synchronous
domain~$\mathrm{SD}$ -- a set of samples of a time series that
satisfy the condition ($\vee$ is the symbol of the logical
operation OR)
\begin{equation} \label{eq:SD}
\begin{aligned}
&\mathrm{SD}_r:\,\left\lbrace
J_{k'} = 1,\; J_{k''} = 0 \vee k'' = 0,\; J_{k'''} = 0 \vee k'''= K + 1
\right\rbrace,\\[3pt]
&k'\in\overline{b^{\mathrm{SD}}_r,\, b^{\mathrm{SD}}_r+L_r-1},\quad
 k''=b^{\mathrm{SD}}_r-1,\quad
 k'''=b^{\mathrm{SD}}_r+L^{\mathrm{SD}}_r,
\end{aligned}
\end{equation}
where~$b^{\mathrm{SD}}_r$, $L^{\mathrm{SD}}_r$, and~$r$ are the
emergence time, the length, and the ordinal number of a
synchronous domain, respectively. In this case, the following
conditions are satisfied: $L^{\mathrm{SD}}_r \leqslant K$, and the
total number of synchronous domains (in the original sequence)
$R^{\mathrm{SD}}\leqslant (K + 1) \operatorname{div} 2$.

To quantitatively describe the structure of synchronization of
systems, the author introduced
in~\cite{bib:article_Makarenko_TechPhysLett_2012_9} the spectral
density function of synchronous domains~$\mathrm{SD}$:
\begin{equation} \label{eq:H_SD_SS}
H^{\mathrm{SD}}\left[L\right] = \sum\limits_{r = 1}^{R^{\mathrm{SD}}} {\delta[L^{\mathrm{SD}}_r,\,{L}]},\quad
L \in \overline {1,\,K},
\end{equation}
where~$\delta[\cdot,\,\cdot]$ is the Kronecker delta.

To analyze the degree of degeneracy of the structure of
synchronous domains, we additionally define a
quantity~$E^{\mathrm{SD}}$ -- the entropy of the structure of
synchronous domains (according to
Shannon)~\cite{bib:article_Makarenko_JETP_2015_5}, which makes
sense for~$\delta^s>0$:
\begin{equation} \label{eq:Entropy_Cond_H_SD}
E^{\mathrm{SD}} = - \sum\limits^K_{i=1}{P^{\mathrm{SD}}\left[i\right]\,\ln P^{\mathrm{SD}}\left[i\right]},\quad
P^{\mathrm{SD}}\left[L\right] =\frac{H^{\mathrm{SD}}\left[L\right]}
{\sum\limits^K_{i=1}{H^{\mathrm{SD}}\left[i\right]}}.
\end{equation}
It follows from Shannon’s entropy properties that the
entropy~$E^{\mathrm{SD}}$ is minimal ($E^{\mathrm{SD}}=0$) when
the spectrum~$H^{\mathrm{SD}}[L]$ is degenerate (all synchronous
domains have the same length) and maximal
($E^{\mathrm{SD}}=\hat{E}^{\mathrm{SD}}$) in the case of a uniform
comb spectrum~$H^{\mathrm{SD}}[L]$ with the maximum number of
different lengths of synchronization domains equal
to~$\hat{W}^{\mathrm{SD}}_{cmb}$:
\begin{equation} \label{eq:Count_differ_L_SD_Max_Entropy_Cond_H_SD}
\hat{W}^{\mathrm{SD}}_{cmb} = \min\left\{
\left\lfloor\frac{\sqrt{1+8\,\delta^s\,K}-1}{2} \right\rfloor,\;
K-\delta^s\,K+1
\right\},\quad
\hat{E}^{\mathrm{SD}} = \ln \hat{W}^{\mathrm{SD}}_{cmb},
 \end{equation}
where~$\lfloor a\rfloor$ is the integer part of~$a$.

On the basis of~(\ref{eq:Entropy_Cond_H_SD}) and~(\ref{eq:Count_differ_L_SD_Max_Entropy_Cond_H_SD}),
we define the relative entropy of the structure of synchronous domains:
\begin{equation} \label{eq:relative_E_Cond_H_SD}
\Delta^{\mathrm{SD}}_E=\frac{E^{\mathrm{SD}}}{\hat{E}^{\mathrm{SD}}}.
\end{equation}
It makes sense to apply the quantity~$\Delta^{\mathrm{SD}}_E$ when
the researcher should compare synchronization cases with different
values of~$\delta^s$ and/or~$K$.

Nevertheless, for the full description of the intermittent
behavior of chaotic systems during synchronization, it is
obviously insufficient to study only synchronous
domains~$\mathrm{SD}$. To obtain a complete and closed idea of the time
structure of synchronism of dynamical systems (a complete and
closed representation of the intermittency structure),
in~\cite{bib:article_Makarenko_JETP_2015_5} the present author
introduced the concept of a {\it desynchronous
domain}~$\mathrm{\overline{S}D}$ -- a set of samples of a time
series satisfying the condition
\begin{equation} \label{eq:NSD}
\begin{aligned}
&\mathrm{\overline{S}D}_r:\,\left\lbrace
J_{k'} = 0,\; J_{k''} = 1 \vee k'' = 0,\; J_{k'''} = 1 \vee k'''= K + 1
\right\rbrace,\\[3pt]
&k'\in\overline{b^{\mathrm{\overline{S}D}}_r,\, b^{\mathrm{\overline{S}D}}_r+L^{\mathrm{\overline{S}D}}_r-1},\quad
 k''=b^{\mathrm{\overline{S}D}}_r-1,\quad
 k'''=b^{\mathrm{\overline{S}D}}_r+L^{\mathrm{\overline{S}D}}_r,
\end{aligned}
\end{equation}
where~$b^{\mathrm{\overline{S}D}}_r$,
$L^{\mathrm{\overline{S}D}}_r$, and~$r$ are the emergence time,
the length, and the ordinal number of a desynchronous
domain~$\mathrm{\overline{S}D}$, respectively.

\begin{figure}
  \centerline{
    \includegraphics[width=0.95\textwidth]{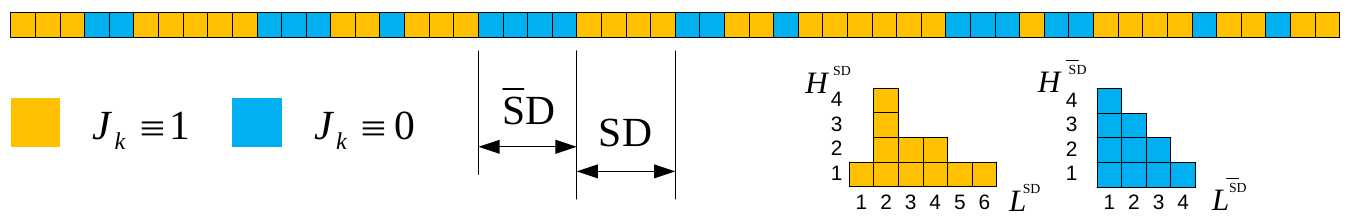}
}
  \caption{Basic characteristics of the time structure of synchronism.}
  \label{fig:TSync}
\end{figure}

The meaning of the characteristics introduced in
this section is demonstrated in Figure~\ref{fig:TSync}
(see~\cite{bib:article_Makarenko_JETP_2015_5} for additional
information).

\subsection{Generalized T-synchronization}

It follows from the definition of the synchronization
condition~(\ref{eq:full_J_T_sync}) that the analyzer proposed
evaluates the complete synchronization
level~\cite{bib:article_Pecora_PhysRevLett_1990_64} and detects
antisynchronization~\cite{bib:article_Liu_PhysLettA_2006_354} with
lag synchronization~\cite{bib:article_Rosenblum_PhysRevLett_1997_78}
precisely in the alphabetic
representation~$\mathrm{T}^{\alpha\varphi}_o$. However, according
to the definition of the geometry of the symbols of the
T-alphabet~(\ref{eq:TSymb}), complete synchronization at the level
of the samples of~$T^{\alpha\varphi}_k$ is a wider phenomenon
compared with the complete synchronization at the level
of~$\mathbf{s}_k$ -- the samples of the sequence itself. The
T-synchronism of dynamical systems (with respect to the set of
phase variables~$\mathbf{s}$) is considered from the viewpoint of
the shape (geometric structure) of the trajectories of the systems
in the extended phase space. By the shape (geometric structure) of
a trajectory of a dynamical system in the extended phase space is
meant its certain invariant under uniform translations and
dilations of the trajectory in the space of phase variables.

Thus, in a sense, the T-synchronization deals with the topological
aspects of synchronization of dynamical
systems~\cite{bib:book_Gilmore_2002, bib:book_Guckenheimer_1997}.
Hence, this opens a possibility for the application of the
analyzer proposed to the study of generalized synchronization of
chaos~\cite{bib:article_Abarbanel_PhysRevE_1996_53}.

To this end, we introduce two additions that relax the
requirements imposed in Section~\ref{sect:Full_T_Sync} on the
complete T-synchronization. The first is the rejection of the
equality of symbols in~(\ref{eq:full_J_T_sync}), and the second is
the rejection of the maximization of~$\delta^s$
in~(\ref{eq:sigma_sync}).

The rejection of the equality of symbols
in~(\ref{eq:full_J_T_sync}) allows one to proceed to the following
definition.

Dynamical systems are \textit{synchronous} at time instant~$k$ in
the sense of Generalized T-synchronization if the
condition~$J_k=1$ is satisfied, where
\begin{equation} \label{eq:gen_J_T_sync}
J_k=\begin{cases}
1 & T^{\alpha\varphi}_k\in\mathrm{M^{JT}},\\
0 & \text{otherwise}.
\end{cases}\;,\quad
\mathrm{M^{JT}}\subseteq\mathrm{T}^{\alpha\varphi}_o|N.
\end{equation}
The structure of the set~$\mathrm{M^{JT}}$ is not trivial. First,
the cardinality of the set is
\begin{equation} \label{eq:card_M_J_T}
\bigl\lvert\mathrm{M^{JT}}\bigr\rvert = \min \Bigl\{
\bigl\lvert\mathrm{V^\Gamma}|_1\bigr\rvert,\,\ldots,\
\bigl\lvert\mathrm{V^\Gamma}|_n\bigr\rvert,\,\ldots,\
\bigl\lvert\mathrm{V^\Gamma}|_N\bigr\rvert\Bigr\}.
\end{equation}
Second, the condition
\begin{equation} \label{eq:unical_symb_M_J_T}
\bigl\lvert\mathrm{M^{JT}}|_{i_n}\bigr\rvert \leqslant 1,\quad
\forall\,\mathtt{T}\,i_n\in\mathrm{V^\Gamma}|_n,\quad
n\in\overline{1,\,N},
\end{equation}
is always satisfied.

As a rule, conditions~(\ref{eq:card_M_J_T}) and
(\ref{eq:unical_symb_M_J_T}) correspond to different variants of
constructing the set~$\mathrm{M^{JT}}$.
Let~$\mathcal{M}^\mathrm{JT}\ni\mathrm{M^{JT}}_j$ be the set of
all admissible variants of constructing the set~$\mathrm{M^{JT}}$,
where~$j$ is the number of a variant.

The number of variants of constructing the set~$\mathrm{M^{JT}}$ is
\begin{equation} \label{eq:Number_M_J_T}
\hat{N}^\mathrm{JT} = \prod\limits^{P_N-1}_{i = 0}
\prod\limits^{N-1}_{n = 1}\Bigl(
\bigl\lvert\mathrm{V^\Gamma}|_n\bigr\rvert-i\Bigr),\quad
P_N = \bigl\lvert\mathrm{V^\Gamma}|_N\bigr\rvert.
\end{equation}
Note that the indices~$n$ in~(\ref{eq:Number_M_J_T}) are
rearranged as follows:
\begin{equation} \label{eq:rear_idx_n}
\bigl\lvert\mathrm{V^\Gamma}|_1\bigr\rvert\geqslant\,\ldots
\geqslant\bigl\lvert\mathrm{V^\Gamma}|_n\bigr\rvert\geqslant\,\ldots
\geqslant\bigl\lvert\mathrm{V^\Gamma}|_N\bigr\rvert.
\end{equation}

If the condition
\begin{equation*} \label{eq:eq_idx_n}
\bigl\lvert\mathrm{V^\Gamma}|_1\bigr\rvert =\,\ldots
=\bigl\lvert\mathrm{V^\Gamma}|_n\bigr\rvert=\,\ldots
=\bigl\lvert\mathrm{V^\Gamma}|_N\bigr\rvert,
\end{equation*}
is valid, then
\begin{equation*} \label{eq:Number_M_J_T_sym}
\hat{N}^\mathrm{JT} = \prod\limits^{P_N-1}_{i = 0}
\Bigl(P_N-i\Bigr)^{N-1}=
\bigl(P_N\bigr)^N\cfrac{\mathrm{\Gamma}^N\bigl(P_N\bigr)}{\mathrm{\Gamma}\bigl(1+P_N\bigr)},
\end{equation*}
where~$\mathrm{\Gamma}$ is the gamma function.

It should also be noted that, in the general
case,~$\hat{N}^\mathrm{JT}$ represents an upper estimate, because
the matrix of~$\Delta^\mathrm{T}$ may contain zero elements; i.e.,
not all possible combinations of elementary symbols are realized.

\begin{figure}
  \centerline{
    \includegraphics[width=168pt]{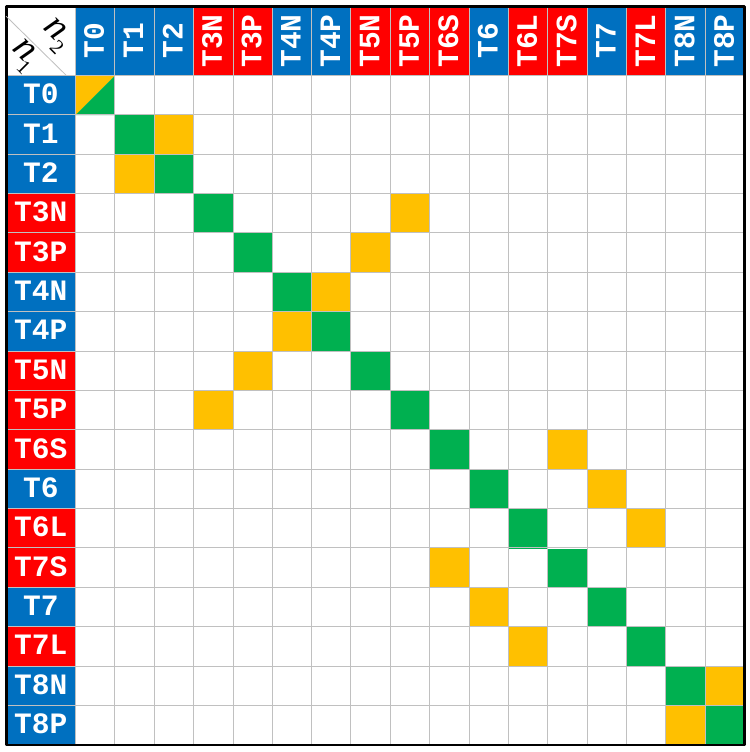}
}
  \caption{The structure of the set~$\mathrm{M^{JT}}$ for~$N=2$ in two cases of
  complete T-synchronization: direct synchronization (green) and antisynchronization (orange).}
  \label{fig:M_JT_full}
\end{figure}

Thus, complete T-synchronization is a special case of generalized
T-synchronization. Figure~\ref{fig:M_JT_full} illustrates the
structure of the set~$\mathrm{M^{JT}}$ for~$N=2$ in two cases of
complete T-synchronization, the direct synchronization and
antisynchronization.

According to~(\ref{eq:Number_M_J_T}), in the case of generalized
T-synchronization, the problem arises of choosing a
set~$\mathrm{M^{JT}}$ from among the
family~$\mathcal{M}^\mathrm{JT}$ that is optimal with respect to
some criterion. In the general case, this is a combinatorial
optimization problem. Let us introduce a generalized algorithm

\begin{equation}\label{eq:alg_M_J_T}
\mathrm{M^{JT}} = \Bigl\{\mathrm{M}^\mathrm{JT}_i:\;F^\mathrm{JT}\to \max,\,
\forall\,\mathrm{M}^\mathrm{JT}_i\in \mathcal{M}^\mathrm{JT} \Bigr\},
\end{equation}
where~$F^\mathrm{JT}$ is a objective function.

For the algorithm~(\ref{eq:alg_M_J_T}), we can consider the
following basic objective functions.

\noindent Maximization of the integral coefficient of synchronism
\begin{equation} \label{eq:Target_funct_0}
F^\mathrm{JT}_{0} = \frac{1}{K}\sum\limits_{i = 1}^K {i\,{H^\mathrm{SD}}\left[ i \right]}  \equiv {\delta ^s}.
\end{equation}

\noindent Maximization of the length of a synchronous domain
\begin{equation} \label{eq:Target_funct_1}
F^\mathrm{JT}_{1} = \max \left\{ {{L^\mathrm{SD}}\left| {{H^\mathrm{SD}}\left[ {{L^\mathrm{SD}}}\right] \ge 1,\;{L^\mathrm{SD}} \in \overline {1,\;K} } \right.} \right\}.
\end{equation}
Note that, if necessary, one can expand the set of objective
functions. Moreover, one can impose additional constraints on
condition~(\ref{eq:alg_M_J_T}).

A naive implementation of the algorithm~(\ref{eq:alg_M_J_T}) leads
to difficulties of computational character. They are associated
with combinatorial explosion. Let us illustrate this situation by
an example of the full T-alphabet,
$\lvert\mathrm{T}^{\alpha\varphi}_o\rvert=17$:

\begin{equation*}
\begin{aligned}
& N = 2, && \hat{N}^\mathrm{JT} = 355\,687\,428\,096\,000,\\[3pt]
& N = 3, && \hat{N}^\mathrm{JT} = 126\,513\,546\,505\,547\,170\,185\,216\,000\,000,\\[3pt]
& N \gg 2, && \text{Curse of dimensionality!}
\end{aligned}
\end{equation*}

Currently, the author has developed a suboptimal version of
algorithm~(\ref{eq:alg_M_J_T}). The main idea of the approach is
as follows: the algorithm operates only with those symbols from
the set~$\mathrm{V^\Gamma}$ for which the elements of the
matrix~$\Delta^\mathrm{T}$ are close to the maximum value. Thus,
this algorithm is free of the need of complete enumeration of the
set~$\mathcal{M}^\mathrm{JT}$; however, it does not guarantee the
global optimum of the function~$F^\mathrm{JT}$. Note that this
topic is the subject of our current research.

\section{Sample}

Let us demonstrate the capabilities of the tools developed by an
example of the analysis of financial time series. The object of
analysis is the time series of exchange rates of some world
currencies (US dollar [USD], Euro [EUR], Japanese Yen [JPH], Swiss
Franc [CHF], and British Pound [GBP] against Russian ruble). The
analyzed period is from~01.01.1999 to~31.12.2014.

\begin{figure}
  \centerline{
    \includegraphics[width=0.95\textwidth]{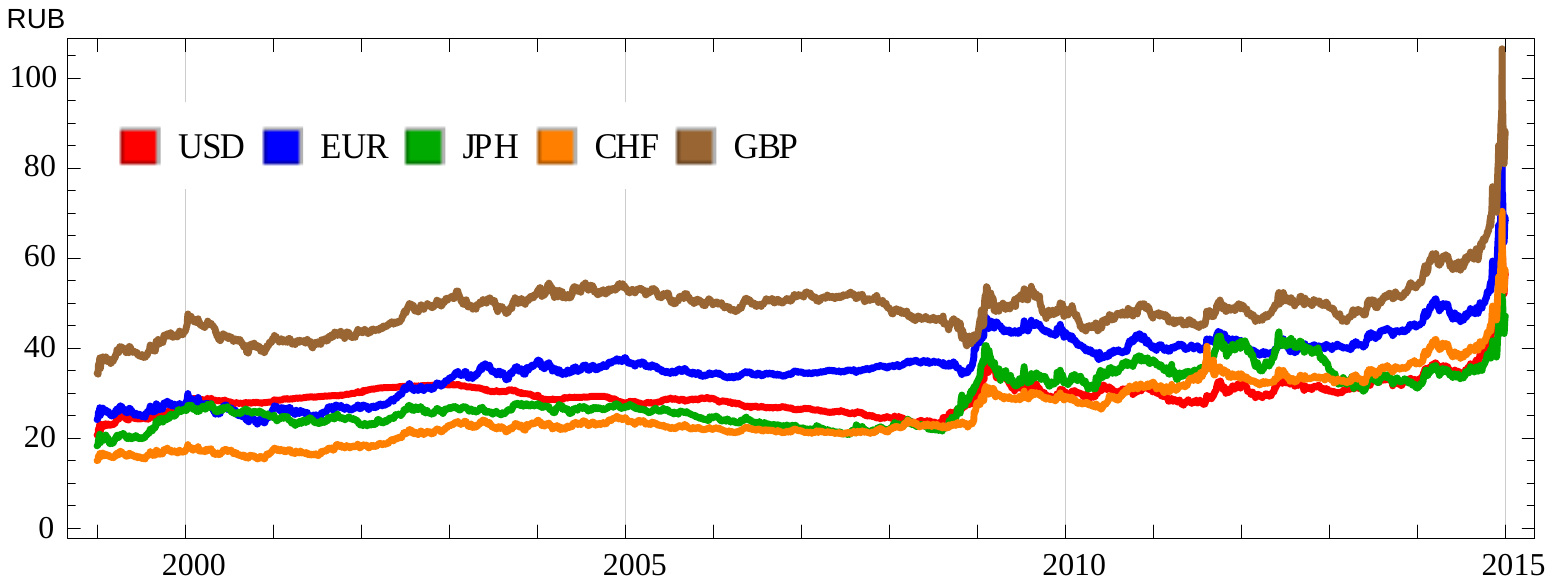}
}
  \caption{Currency exchange rates.}
  \label{fig:Curency_Exch}
\end{figure}

Note that the analysis of the Generalized T-synchronization has
applied value in the context of research in macroeconomics and
stochastic financial mathematics. The original data are taken from
the official web-site of the Central Bank of Russia (Bank of
Russia, exchange rates, www.cbr.ru/eng/). The length of the time
series is $K=3\,985$~samples. The initial time series are shown in
Figure~\ref{fig:Curency_Exch}. Note that these time series have also been studied from the
viewpoint of complete T-synchronization~\cite{bib:report_Makarenko_AFS_2013} (the
analyzed period is from~01.01.1999 to~31.03.2013) and
TQ-complexity~\cite{bib:report_Makarenko_MICNON_2015}.

We carried out a detailed analysis of generalized
T-synchronization for the USD/EUR pair. For comparison, we also
evaluated the characteristics of complete synchronization. The
set~$\mathrm{M^{JT}}$ was constructed for the objective
function~$F^\mathrm{JT}_{0}$. As a result, we obtained the
following values of the integral coefficient of synchronism:
\begin{equation} \label{eq:delta_s_Curency_Exch}
\delta^s|C = 0.174492,\quad
\delta^s|A = 0.219433,\quad
\delta^s|G = 0.222948,
\end{equation}
where~$\delta^s|C$, $\delta^s|A$, and~$\delta^s|G$ are complete,
anti-, and generalized regimes of synchronization, respectively.

\begin{figure}
  \centerline{
    \includegraphics[width=0.5\textwidth]{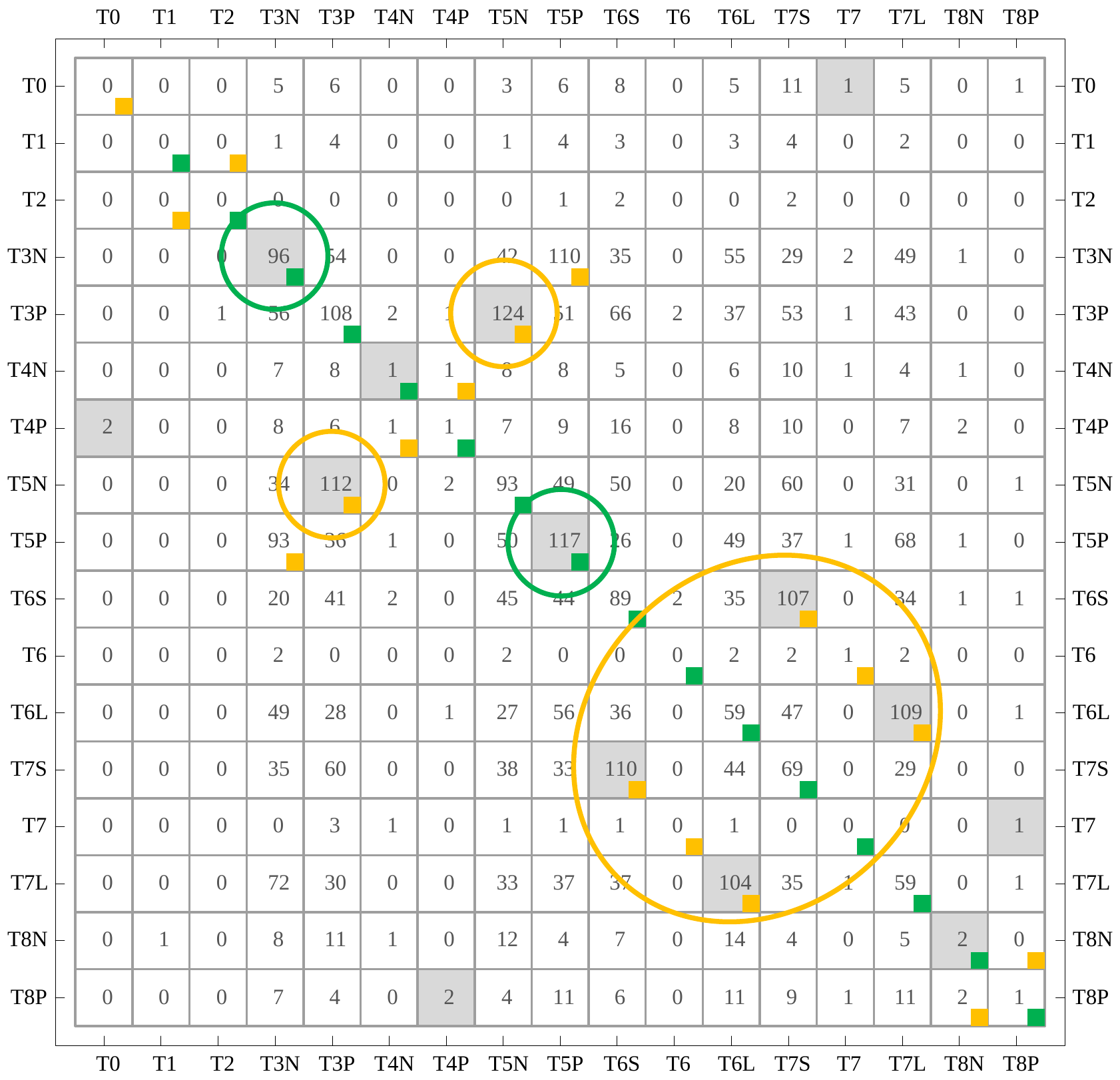}
}
  \caption{Matrix~$\Delta^T$; grey cells are T-symbols included in the set~$\mathrm{M^{JT}}$;
   green and orange squares demonstrate the structure of the set~$\mathrm{M^{JT}}$ in two cases of
   complete T-synchronization (see Figure~\ref{fig:M_JT_full}); green and orange ellipses are key
   T-symbols in the set~$\mathrm{M^{JT}}$.}
  \label{fig:Curency_Exch_M_JT}
\end{figure}

The results~(\ref{eq:delta_s_Curency_Exch}) show that the regime
of generalized T-synchronization is characterized by the maximum
value of the integral coefficient of synchronism. At the same
time, from the structural point of view, this mode is
characterized by the combination of direct- and anti-synchronism
regimes (see Figure~\ref{fig:Curency_Exch_M_JT}).

\begin{figure}
  \centerline{
    \includegraphics[width=0.8\textwidth]{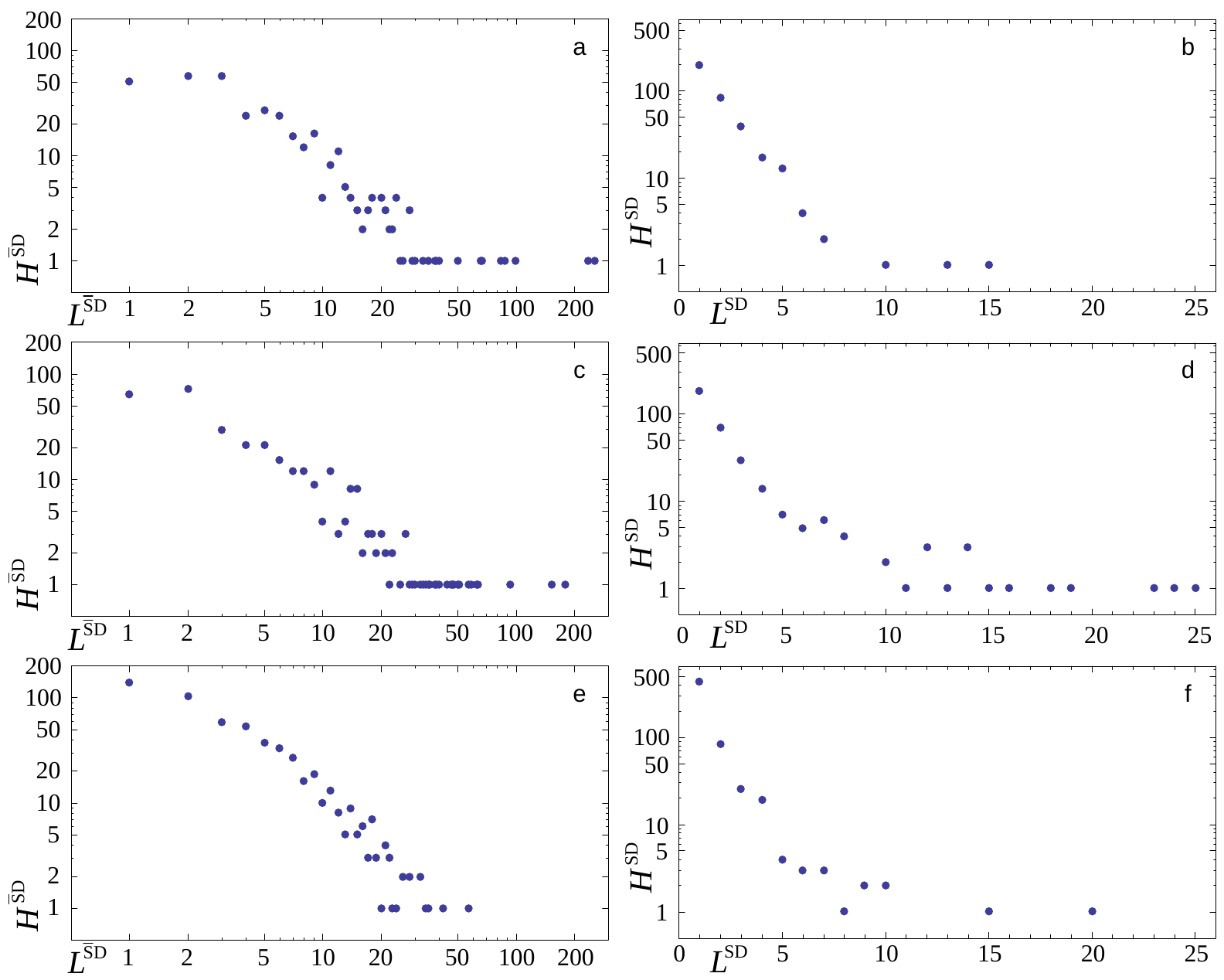}
}
  \caption{Spectral density of domains~$\mathrm{\overline{S}D}$ and~$\mathrm{SD}$;
  (a, b), (c, d), and~(e, f) are complete, anti-, and generalized regimes of synchronization,
  respectively.}
  \label{fig:Curency_Exch_HSS}
\end{figure}

Taking account of the factor of generalized synchronization for
the pair~USD/EUR modifies the time structure of its
T-synchronism (see Figure~\ref{fig:Curency_Exch_HSS}).

Figure~\ref{fig:Curency_Exch_HSS} shows that taking account of the
factor of generalized synchronization significantly reduces the
maximum length of desynchronous domains. In this case, the mean
length of synchronous domains also decreases.

\section{Conclusion}

We have proposed a new method to diagnose generalized
synchronization in nonlinear multidimensional chaotic systems. The
method allows one to explore and quantitatively evaluate the
time structure of synchronization of chaotic oscillations in
the so-called T-synchronization
regime~\cite{bib:article_Makarenko_TechPhysLett_2012_9,
bib:article_Makarenko_JETP_2015_5}. The approach is based on the
formalism of symbolic CTQ-analysis, which was proposed by the
author in~\cite{bib:article_Makarenko_TechPhysLett_2012_38_155,
bib:article_Makarenko_ComputMathMathPhys_2012_52_1017}.

The method considered, which is based on the analysis of
generalized T-synchronization, can be successfully applied to the
study of multidimensional systems consisting of two or a greater
number of coupled nonidentical oscillators, including
multidimensional lattices of oscillators with arbitrary topology.
The approach described can be applied to the analysis of
experimental data, because it does not require any a priori
knowledge of the system under study.

\begin{Biblioen}

\bibitem{bib:book_Pikovsky_2001}
A.S.~Pikovsky, M.G.~Rosenblum, and J.~Kurths. {\em
Synchronization: A universal concept in nonlinear sciences.}
Cambridge University Press, Cambridge, 2001.

\bibitem{bib:article_Boccaletti_PhysRep_2002_366}
S.~Boccaletti, J.~Kurths, G.V.~Osipov, D.L.~Valladares, and C.S.~Zhou. {\em The synchronization of chaotic systems.} Physics Reports {\bf 366}, 1, 1--101, 2002.

\bibitem{bib:article_Argonov_JETPL_2004_80}
V.Yu.~Argonov and S.V.~Prants. {\em Synchronization and bifurcations of internal and external degrees of freedom of an atom in a standing light wave.} J.~Exp.~Theor.~Phys.~Lett. {\bf 80}, 4, 231--235, 2004.

\bibitem{bib:article_Kuznetsov_UFN_2011_2}
S.P.~Kuznetsov. {\em Dynamical chaos and uniformly hyperbolic
attractors: From mathematics to physics.} Phys.~Usp. {\bf 54}, 2,
119--144, 2011.

\bibitem{bib:article_Napartovich_JETP_1999_5}
A.P.~Napartovich and A.G.~Sukharev.{\em Synchronizing a chaotic laser by injecting a chaotic signal with a frequency offset.} J.~Exp.~Theor.~Phys. {\bf 88}, 5, 875--881, 1999.

\bibitem{bib:article_Cuomo_PhysRevLett_1993_71}
K.M.~Cuomo and A.V.~Oppenheim. {\em Circuit implementation of synchronized chaos with applications to communications.} Phys.~Rev.~Lett. {\bf 71}, 1, 65--68, 1993.

\bibitem{bib:article_Larger_Physique_2004_5}
L.~Larger and J.-P.~Goedgebuer. {\em Encryption using chaotic dynamics for optical telecommunications.} C.R.~Physique {\bf 5}, 6, 609--611, 2004.

\bibitem{bib:article_Planat_Neuroquantology_2004_2}
M.~Planat. {\em On the cyclotomic quantum algebra of time
perception.} Neuroquantology {\bf 2}, 4, 292--308, 2004;
arXiv~quant-ph/0403020.

\bibitem{bib:article_Abarbanel_PhysRevE_1996_53}
H.D.I.~Abarbanel, N.F.~Rulkov, and M.M.~Sushchik. {\em Generalized synchronization of chaos: The auxiliary system approach.} Phys.~Rev.~E {\bf 53}, 5, 4528, 1996.

\bibitem{bib:article_Pecora_PhysRevLett_1990_64}
L.M.~Pecora and T.L.~Caroll. {\em Synchronization in chaotic systems.} Phys.~Rev.~Lett. {\bf 64}, 8, 821--824, 1990.

\bibitem{bib:article_Liu_PhysLettA_2006_354}
W.~Liu, X.~Qian, J.~Yang, and J.~Xiao. {\em Antisynchronization in coupled chaotic oscillators.} Phys.~Lett.~A {\bf 354}, 1--2, 119--125, 2006.

\bibitem{bib:article_Rosenblum_PhysRevLett_1997_78}
M.G.~Rosenblum, A.S.~Pikovsky, and J.~Kurths. {\em From phase to
lag synchronization in coupled chaotic oscillators}
Phys.~Rev.~Lett. {\bf 78}, 22, 4193, 1997.

\bibitem{bib:article_Anishenko_TechPhysLett_1988_6}
V.S.~Anishchenko and D.E.~Postnov. {\em Effect of the locking of the basic frequency for chaotic self-oscillations. Synchronization of strange attractors.} Sov.~Tech.~Phys.~Lett. {\bf 14}, 3, 254--258, 1988.

\bibitem{bib:article_Pikovsky_JourBifChaos_2000_10}
A.S.~Pikovsky, M.G.~Rosenblum, and J.~Kurths. {\em Phase synchronization in regular and chaotic systems.} Int.~J. of Bifurcation and Chaos. {\bf 10}, 10, 2291, 2000.

\bibitem{bib:article_Koronovskii_JETPL_2004_79}
A.A.~Koronovskii and A.E.~Khramov. {\em Wavelet transform analysis of the chaotic synchronization of dynamical systems} J.~Exp.~Theor.~Phys.~Lett. {\bf 79}, 7, 316--319, 2004.

\bibitem{bib:article_Makarenko_JETP_2015_5}
A.~Makarenko. {\em Analysis of the time structure of synchronization in multidimensional chaotic systems.}  J.~Exp.~Theor.~Phys. {\bf 120}, 5, 912--921, 2015; arXiv:~1505.04314.

\bibitem{bib:article_Makarenko_TechPhysLett_2012_9}
A.V.~Makarenko. {\em Measure of synchronism of multidimensional chaotic sequences based on their symbolic representation in a T-alphabet.} Tech.~Phys.~Lett. {\bf 38}, 9, 804--808, 2012; arXiv:1212.2724.

\bibitem{bib:article_Makarenko_TechPhysLett_2012_38_155}
A.V.~Makarenko. {\em Structure of synchronized chaos studied by symbolic analysis in velocity-curvature space.} Tech.~Phys.~Let. {\bf 38}, 2, 155--159, 2012; arXiv:1203.4214.

\bibitem{bib:article_Makarenko_ComputMathMathPhys_2012_52_1017}
A.V.~Makarenko. {\em Multidimensional dynamic processes studied by
symbolic analysis in velocity-curvature space.} Comput.~Math.
and~Math.~Phys. {\bf 52}, 7, 1017--1028, 2012.

\bibitem{bib:book_Gilmore_2002}
R.~Gilmore and M.~Lefranc. {\em The topology of chaos.} Wiley-Interscience, New York, 2002.

\bibitem{bib:book_Guckenheimer_1997}
J.~Guckenheimer and P.~Holmes. {\em Nonlinear Oscillations, Dynamical Systems and Bifurcation of Vector Fields.} Springer, New-York, 1997.

\bibitem{bib:article_Bowen_AJM_1973_95}
R.~Bowen. {\em Symbolic dynamics for hyperbolic flows.} Amer.~J.~Math. {\bf 95}, 429--460, 1973.

\bibitem{bib:report_Makarenko_CHAOS_2015}
A.V.~Makarenko. Generalized synchronization of multidimensional chaotic systems in terms of symbolic CTQ-analysis. {\em Book of Abstracts of the 8th Chaotic Modeling and Simulation International Conference (CHAOS~2015).} Paris, ISAST, IHP, 2015, pp.~77--78.

\bibitem{bib:article_Zeldovich_UFN_1987_5}
Ya.B.~Zel’dovich, S.A.~Molchanov, A.A.~Ruzmaikin, and D.D.~Sokolov. {\em Intermittency in random media.} Sov.~Phys.~Usp. {\bf 30}, 5, 353--369, 1987.

\bibitem{bib:article_Mandelbrot_JFluidMech_1974_2}
B.B.~Mandelbrot. {\em Intermittent turbulence in self-similar
cascades: Divergence of high moments and dimension of the carrier}
J.~Fluid~Mech. {\bf 62}, 2, 331--358, 1974.

\bibitem{bib:report_Makarenko_AFS_2013}
A.V.~Makarenko Symbolic CTQ-analysis -- a new method for studying of financial indicators. {\em Book of Abstract of the International Conference "Advanced Finance and Stochastics" (AFS~2013)}. Moscow, Steklov Mathematical Institute, 2013. P.~63.

\bibitem{bib:report_Makarenko_MICNON_2015}
A.V.~Makarenko. Estimation of the TQ-complexity of chaotic sequences. {\em Proceedings of the 1st IFAC Conference on Modelling, Identification and Control of Nonlinear Systems (MICNON~2015)}. Saint Petersburg, IFAC, 2015; arXiv:~1506.09103.

\end{Biblioen}


\noindent
\\\textsf{\textbf{Andrey V. Makarenko} -- was born in~1977, since~2002 -- Ph.~D. of Cybernetics. Founder and leader of the Research \& Development group "Constructive Cybernetics". Author and coauthor of more than 60~scientific articles and reports. Member~IEEE (IEEE Signal Processing Society Membership; IEEE Computational Intelligence Society Membership). Research interests: Analysis of the structure dynamic processes, predictability; Detection, classification and diagnosis is not fully observed objects (patterns); Synchronization and self-organization in nonlinear and chaotic systems; System analysis and math.~modeling of economic, financial, social and bio-physical systems and processes; Convergence of Data~Science, Nonlinear~Dynamics, and~Network-Centric.}

\end{document}